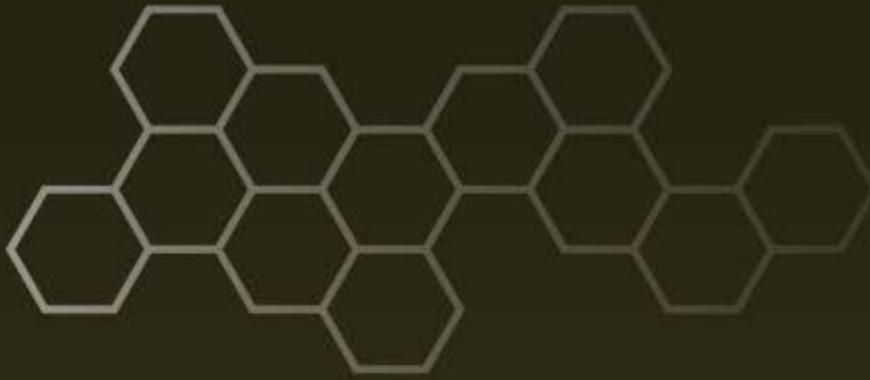

ARL-TR-7566 ● DEC 2015

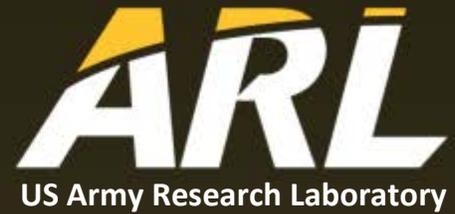

US Army Research Laboratory

# Assessing Mission Impact of Cyberattacks: Report of the NATO IST-128 Workshop

by Alexander Kott, Nikolai Stoianov, Nazife Baykal, Alfred Moller, Reginald Sawilla, Pram Jain, Mona Lange, and Cristian Vidu

**NOTICES**

**Disclaimers**

The findings in this report are not to be construed as an official Department of the Army position unless so designated by other authorized documents.

Citation of manufacturer's or trade names does not constitute an official endorsement or approval of the use thereof.

Destroy this report when it is no longer needed. Do not return it to the originator.

ARL-TR-7566 ● DEC 2015ARL-TR-7566 ● DEC 2015

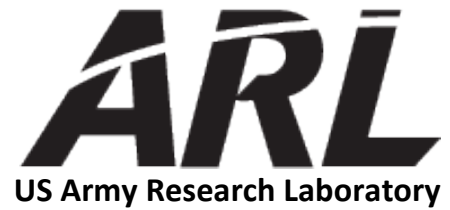

US Army Research Laboratory

# Assessing Mission Impact of Cyberattacks: Report of the NATO IST-128 Workshop

by Alexander Kott, *Computational and Information Sciences Directorate, ARL*

Nikolai Stoyanov, *Defense Institute, Bulgaria*

Nazife Baykal, *Middle Eastern Technical University, Turkey*

Alfred Moller, *Danish Defence Acquisition and Logistics Organization, Denmark*

Reginald Sawilla, *NATO Communications and Information Agency, Netherlands*

Pram Jain, *MITRE, USA*

Mona Lange, *University of Lübeck, Germany*

Cristian Vidu, *National University for Political Sciences and Public Administration, Romania*
Approved for public release; distribution unlimited.

# Contents

## Acknowledgments

We wish to acknowledge many people and organizations who made this workshop a success. The Istanbul Technical University provided facilities for the workshop. Mr. Oğuzhan Topgül and Mr. Hüseyin TİRLİ, researchers of the TÜBİTAK BİLGEM Cyber Security Institute, Turkey, tirelessly supported the workshop before and during its time in session. Dr. John McLean, the Chairman of the Information Systems and Technology (IST) panel, encouraged creation of this workshop and personally participated in the proceedings. Mrs. Aysegül Apaydin, IST panel assistant, guided, advised and managed numerous programmatic and logistical aspects of the workshop. Ms. Ana Santiesteban, US Army Research Laboratory, helped to manage the flow and review of the papers submitted to the workshop. The members of the workshop's technical committee (see Appendix B for the list of names) generously provided their time to formulate the workshop's program and to review the papers.

## 1. Introduction

This report describes the proceedings and outcomes of the North Atlantic Treaty Organization (NATO) workshop IST-128 / RWS-019, titled "Cyber Attack Detection, Forensics and Attribution for Assessment of Mission Impact," organized by the NATO Science and Technology Organizations' (STO) Information Systems and Technology (IST) panel. The workshop was held at the Istanbul Technical University, in Istanbul, Turkey, on 15–17 June 2015. The workshop was NATO Unclassified and open to the NATO Partnership for Peace (PfP) nations, and to the NATO Mediterranean Dialogue (MD) nations.

The STO's mission is to help position the Nations' and NATO's science and technology (S&T) investments as a strategic enabler of the knowledge and technology advantage for the defense and security posture of NATO Nations and partner Nations. This is accomplished by conducting and promoting S&T activities that augment and leverage the capabilities and programs of the Alliance, of the NATO Nations, and the partner Nations, in support of NATO's objectives. It is further accomplished by contributing to NATO's ability to enable and influence security and defense-related capability development and threat mitigation in NATO Nations and partner Nations, in accordance with NATO policies; and by supporting decision making in the NATO Nations and NATO.

The immediate sponsor of this workshop, IST is one of the 7 panels whose role it is to implement, on behalf of the S&T Board, the STO mission with respect to information systems technology. The advancement and exchange of techniques and technologies to provide timely, affordable, dependable, secure, and relevant information to warfighters, planners, and strategists, as well as enabling technologies for modelling, simulation, and training are the focus of this panel. The IST covers the fields of information warfare and assurance, information and knowledge management, communications and networks and architecture and enabling technologies.

The motivation for the workshop had to do with the fact that the success of a military mission is highly dependent on the communications and information systems (CISs) that support the mission and their use in the cyber battlespace. The inexorably growing dependency on computational information processing for weapons, intelligence, communication, and logistics systems continues to increase the vulnerability of missions to various cyber threats. Attacks on CISs or other cyber incidents degrade or disrupt the usage of CISs, and the resulting mission capability, performance, and completion. Such incidents are expected to increase in frequency and sophistication.

Thus, in initiating this workshop, the IST panel felt that there is a need to address the technology and procedures to characterize the impact of cyber-attacks on the mission. Such an impact analysis must necessarily include a broad range of cyber analysis activities: detect attacks in a mission-supporting manner, assess damages relevant to the mission, investigate impacts on mission elements, recover from attacks in order to continue missions to the maximum extent possible, and decide on how to respond to cyberattacks in a manner that maximizes mission success. Additionally, the IST panel believed, forensics methods and tools are necessary to determine key facts relevant to assessing mission impact. Such tools are used for evidence collection, analysis of the attack, identification of the attacker, understanding the attack, damage assessment, and attribution of attackers. Dependent on the mission and the type of an attack, there may be different degrees of relative importance and resources attached to attack detection, continuity of the military mission, damage assessment, evidence collection, attribution, and other activities. Usage of related methods, procedures, tools, or technology will depend on the requirements of the mission.

This workshop, therefore, was to focus on identifying practice and research challenges, gaps, and approaches – current and future – to assessment of mission impact due to a cyberattack. Because it was thoughts that such an assessment is inseparable from, and impossible without, attack detection, forensics, and attribution, the workshop also intended to explore how these activities and related technologies and methods should support the assessment of mission impact. The initial announcement of the workshop is found in Appendix A. The final program of the workshop is found in Appendix B.

The workshop gathered a total of 56 participants from 9 countries. The participants included authors and presenters of 12 technical papers and briefs. The workshop attracted 2 types of papers: full papers and position papers. A full paper is a technical paper (between 2500–5000 words) that presents results of novel research and is subject to evaluation criteria of a typical technical conference. A position paper is short (between 500–1500 words) and reflects the views of the author – usually a discussion of research challenges, gaps, and approaches – without necessarily presenting a supporting research. The papers are found in an accompanying volume: *Proceedings of the NATO IST-128 Workshop: Assessing Mission Impact of Cyberattacks*.[1]

The introductory session of the workshop comprised 3 talks that welcomed the participants and offered an overview of the organizations and topics involved. It was followed by 4 technical sessions that totaled 12 technical talks.

The first technical session focused on the need to gain insights into the intent, motivations, and capabilities of the attackers, in order to understand the intended and actual mission impact. The second technical session explored whether, and to what extent, it is possible to understand the mission impact by analyzing the observable cyber signal and events, through such means as are normally associated with cyber intrusion detection, forensics, and malware analysis. The third technical session discussed the need for models of missions and systems that support missions and the approaches to constructing such models. The fourth technical session investigated the means by which mission impact could be simulated or modeled.

The last day of the workshop was limited to 1 half-day session. In a structured brainstorming process, that session integrated and analyzed the overall results and findings of the workshop. The participants divided into 3 brainstorming teams. The first team aimed to summarize the state of the art in mission impact assessment, including the key gaps. The second team was asked to envision the future state of the mission impact assessment 10–20 years from now. The third analyzed the discussions of the 4 previous sessions in order to distill the key research directions that the community should pursue. The 3 teams then presented and discussed their findings in a plenary session. Finally, the workshop formulated key ideas for the follow-on activity for consideration by the NATO IST panel.

A few disclaimers are in order here. First, not every author of the report or participant of the workshop agrees with every (or any) opinion presented in the workshop's report. Second, all statements of fact or opinion presented in this report are those of the workshop participants and do not reflect positions or views of their employers or any organizations with which they are affiliated.

## 2.   Findings of the Workshop

### 2.1  The Key Finding: The Primacy of a Model-driven Paradigm

The original expectations of the workshop's organizers, although not particularly explicit, were reflected in the title of the workshop: it was assumed that the key to addressing the challenges of mission impact assessment had to do with a form of integration of intrusion detection, analysis, forensics, attribution, and related traditional disciplines of cyber defense. The answer that emerged from the workshop, however, was decidedly different.

The workshop witnessed the emergence of a very strong and clear consensus. The key to solving the mission impact assessment problem, the workshop's participants argued, was in adopting and developing a new model-driven paradigm. It requires creation and validation of mechanisms of modeling the organization whose mission is subject to assessment, the mission (or missions) itself, and the cyber-vulnerable systems that support the mission. The models are then used to simulate or otherwise portray the impacts of the cyber-attacks.

In addition, such model-based analysis could be used to explore multiple alternative mitigation and work-around strategies – an essential part of coping with mission impact – and select the optimal course of mitigating actions. Only such a paradigm can be expected to provide meaningful, actionable information about the mission impacts that have not been seen before or do not match prior experiences and patterns.

To be sure, the model-driven paradigm of mission impact assessment does not imply that traditional disciplines of cyber defense, such as vulnerability analysis, intrusion prevention, intrusion detection, analysis, forensics, attribution, and recovery, are irrelevant to the topic of impact assessment. Rather, these tasks themselves can benefit from – and be integrated into the overall framework of – the model-driven paradigm. For example, intrusion detection, especially for zero-day or polymorphic attacks, would greatly benefit from the ability to model the observable effects of a hypothetic attack.

Based on these findings, the workshop proposed an exploratory team activity that focuses on a study of potential approaches and likely barriers to applying model-driven paradigm to a broad range of cyber defense activities, including but not limited to mission impact assessment.

## 2.2 Mission Impact Assessment Problem Formulation

Appropriate formulation of a problem is the key to its successful solution. What constitutes a successful mission impact assessment (MIA) solution depends, in turn, on who are the users of the solution. Therefore, the workshop's participants argued for more involvement of the "stakeholders" – primarily end users – in formulating the MIA problem. For example, commanders need decision support at "commander" level – they are much less interested in technical details. For these reasons, future techniques of MIA may specifically focus – and the very problem of MIA may be so formulated – on supporting the cyber security decision-making process, and particularly on tools that teach and train decision makers, and support them.

Determining the correct users, however, depends on knowing where MIA belongs in the broader scheme of things. One way to position MIA in the context of a

Approved for public release; distribution unlimited.

broader cyber defense concept is to consider MIA as a part of the big control loop that strives to keep the controlled "plant" – the mission – within the prescribed space of secure states. The output of this controller is a set of corrective actions, or a course of actions, designed to keep the plant in the secure state. Parenthetically, among the actions that could be taken to correct or prevent the mission impact one is particularly salient – deterrence. MIA can help formulate the nature and extent of the appropriate deterrence action.

In this formulation, then, MIA is the component of the control system that measures how much the plant has deviated or will deviate from the desired state. Once we say "how much," we must consider what exactly we quantify when we measure the deviation from security to lesser security. To put it differently, a formal quantification of a utility function is needed. Some of the current approaches to MIA are based on heuristic scoring. To oversimplify, the assessor sums up the "impact points" and declares that the total impact on the mission is, let's say, 73 points. Clearly, this leaves much to be desired in terms of theoretical rigor. Similarly, to say that "The mission impact is 70% failure" is very difficult to interpret. For example, even with a 70% mission failure (whatever that means), the operator may still be able to reach a key goal.

Perhaps a more principled approach to measures would be to use traditional characteristics of security – confidentiality, integrity, and availability (CIA). Unfortunately, aspects of CIA for missions are not well defined or understood – in particular, measuring integrity and confidentiality is poorly understood. Furthermore, one can argue that in cyber-physical systems CIA is not particularly meaningful, and a better triad is observability, controllability, and operability (OCO).

Other ways to express quantitative output of MIA would be to measure mission impact as a reduction in tangible attributes of the system, such as the network bandwidth, delay, or power use. Yet another approach would be to quantify the distance from the achievable states to desired states, for example, via the cost of the corrective actions that would bring the plant to the desired, secure state. All this suggests that a formal language, a formal mathematics of mission security, would be highly desirable to give MIA a solid quantitative foundation.

Appropriate formulation of the MIA problem also requires a choice of the right level of abstraction. When formulated and solved at a very abstract level, the solution may not give adequate insights into what actions – often very specific and detailed – need to be considered. On the other hand, when formulated at a very detailed level, the problem demands a very intricate model that is far too expensive to construct. Some workshop participants argued that a shift must occur from the enterprise-scale problem – common in today's practice – to more meaningful tactical scale. One argument for more detailed formulations is that that seemingly small attacks on mission activities can have large effects, as confirmed by some simulation studies.

Although so far we discussed mainly the control-theoretic style of the MIA problem formulation, we should not overlook the game-theoretic (or game-playing simulation) perspective. This would be natural, considering that the cybersecurity problem is highly adversarial in nature. Because it involves intelligent, strategically thinking adversaries, as opposed to random deviations in the plant operation, a game perspective seems more appropriate than a control perspective. A related and appropriate style of problem formulation could be a robust control with adversarial inputs. One could also consider a contest-game theory framework for modelling economic impacts of cyberattacks. Because full information is not normally available, the problem should be formulated as a partial information game; artificial intelligence techniques might help here.

Yet another style of problem formulation that might be applicable to MIA is the risk analysis problem. In fact, risk analysis and MIA are closely related. Arguably, instead of saying, "Impact to Mission M from attack A is X," one could say nearly equivalently, "Risk to Mission M from attack A is Y." Furthermore, risk is successfully used in ecological impact studies, and an ecological (and biological) analogy may be instructive for complex cyber systems. The risk flow concept used in analysis of ecological systems appears to be a promising cybersecurity approach that may contribute to a temporal and spatial assessment of risk propagation and influence. In addition, many risk analysis techniques do not take into account a strategically thinking adversary, as game formulations do.

While arguing for a model-driven paradigms for MIA, with either a control, game, or risk perspective, we must not dismiss entirely different yet complementary approaches. For example, analysis of malware found on friendly systems or otherwise captured in the wild can be an important help in MIA. Malware analysis and reverse engineering may reveal the adversary's intent regarding the mission impact that the malware was designed to inflict.

If all the above sounds very complex, that is because it is. Despairing of this complexity, one might be tempted to ask if a much simpler formulation might be adopted. For example, a pattern-matching approach like, "if you see pattern of evidence E, conclude that impact is X." However, the workshop's participants did not feel that the MIA problem lends itself to much simplification. Instead, model-driven approaches were deemed most promising, which brings us to the question of what should be included in a good model.

## 2.3 Model Content

The workshop exhibited a rather strong agreement on the fundamental components of a model required for MIA. These were generally understood to be the models of the organization (or military unit), its business processes (often decomposed into functions and tasks), the missions executed through the business processes, and the technical systems that support the missions. Relations, influences, and dependencies – quantitatively characterized – between all these entities and their sub-entities need to be modeled. Even the physical environment of a mission may need to be modeled, as well as the sensors and actuators that sense and affect the environment, because they also can be subjects of cyber or deception attacks.

Because the MIA problem is so fundamentally adversarial in nature, it was widely recognized that one needs a comprehensive model for adversary characterization, and behavior understanding and prediction. Models should also include an environment property, attacks property, and target property, including modeling of these 3 elements and relations and interactions between them. We discuss adversary modeling in more detail in the next section.

In addition to describing the structure of the problem, models must capture its dynamics. There are several very different meanings of dynamics in MIA models. First, the structure itself changes rapidly. For example, the servers supporting a mission might be taken down for maintenance and then brought up on line again or reassigned to another mission. The model would need to be updated continually to reflect such changes. Second, when a cyber-attack impacts a mission, the defenders and operators of the mission and supporting systems often show remarkable ability to work around an established process, for example, to redesign the business process rapidly and radically. Third, even in a very static structure of the business, actions are dynamic – they start, proceed, and stop at points in time. This dynamic has to be captured in a model. Fourth, the characteristics of components and relations within the model may change depending on the context. For example, the criticality of systems change during different missions, various simultaneous missions can each place a different criticality on the same shared systems, and the organization will have a time-dependent and dynamically changing aggregate dependency on its systems and other assets.

Dynamics associated with the adversary's behavior are particularly complex, and we consider them next.

## 2.4 Models of Adversary

Mission impact has to be considered in the context of what impact the adversary desires. If we know, or are able to estimate, the intents, motivations, and anticipations of the adversary, the impact of the adversary on our missions, or the intended impact, would be easier to assess. It should be noted that in this section we consider the adversary rather abstractly; in particular, we do not assume that the adversary can be modeled as an individual human or a collection of individual humans. That perspective is explored in a later section.

A model for adversary characterization and behavior understanding and prediction should be sufficiently comprehensive. In particular, the model should include properties of the environment in which the cyber conflict occurs; the properties of the attacks and targets that area available to the adversary; and relations and interactions between all such elements, all this in addition to the properties of the adversary itself.

Naturally, properties and characteristics of the adversary are often unknown, or can be assumed only with a significant degree of uncertainty. Modeling tools should allow representation of such uncertainties and even unknowns.

Modeling of the cyber adversary can be helped by analysis of its testing of cyber weapons, of which a number of examples are known. Testing of weapons is critical to the adversary and is detectable by us. Tests tell us much about the adversary and about potential mission impacts. Therefore, we need a program, framework, and sensors to be developed for collection and analysis of such tests.

We can also learn from modeling experiences in earlier decades and against different types of weapons. In particular, work done on understanding the adversary behaviors for purposes of nuclear exchanges has been seen as fairly successful (for example, no nuclear exchange has happened to this date) and is potentially a useful analogy for cyber conflicts. On the other hand, how sure are we that we really know interests, intentions, and strategies of nuclear adversaries? The Cuban crisis seems a counter-example. Is it not a misleading analogy? For example, a cyber conflict is much less constrained in its effects than a nuclear conflict.

A powerful determinant of adversary behavior is the adversary's expectations of our response. Thus, it is important to understand the role of deterrence – the measures we can take to prevent hostile actions by an adversary – in a cyber conflict. An adversary model should help answer questions like, what does the adversary wants to do and what do they expect us to do?

Currently, little is either understood or practiced with respect to deterrence. Arguably, with a consistent and well-understood practice of deterrence, impacts of cyber actions by an adversary may be more predictable. Unlike in the case of nuclear deterrence, cyber deterrence might be possible without a threat of human victims; the effects could be strictly limited to impacts on industrial and financial infrastructure, for example.

Yet another difference between cyber and nuclear cases is that the theory of nuclear conflict relies on a geopolitical approach. Cyber operations, in the other hand, are largely geography-independent, difficult to attribute to a specific state, and often perpetrated by non-state actors, or state-sponsored but otherwise non-state actors. Economic rather than geopolitical perspectives might work better in the case of a cyber conflict.

Analogies other than nuclear conflict might be better for cyber conflicts. The workshop participants discussed the possible terrorism and counter-terrorism analogies. Whether we can claim much success in understanding and modeling terrorist adversaries is an open question.

## 2.5 Models of Humans

When the adversary is an individual human, or a group of individuals that we find appropriate to model individually, we should consider techniques of cognitive modeling of individual human minds. Such models can help predict how adversaries (cyber-attackers) formulate their goals and thereby tell us about the intended or actual mission impact of the adversary's actions. Tools like Adaptive Control of Thought—Rational (ACT-R) are popular for cognitive modeling and might be applicable to modeling behaviors of cyber-attackers. Cognitive modeling of individuals is proven to be possible, and validated tools for such modeling do exist. Still, this area tends to be a rather early research field.

Although we have focused on modeling the adversary, models of defenders should not be overlooked. To assess the likely mission impact, we need to know how a human cyber defender reacts to cyber-attacks. Errors committed by defenders determine the extent of mission impact. A defender may fail to recognize a threat and to take appropriate actions, thereby enabling a greater mission impact. A defender may fall a victim to deception committed by an attacker. A defender may fail to undertake a suitable work-around when a mission is impacted. A defender may also misinterpret mission impacts when they occur. All this is highly relevant to the MIA problem.

Whether one models an attacker or a defender, the model needs to be rich enough to reflect "irrational" aspects of human cognition, such as cognitive biases. These are particularly important in the high-pressure, high-tempo, non-intuitive world of cyber operations. Impact of dynamic learning must be considered to account for rapid evolution of knowledge in cyber conflicts. Game-theoretic approaches should be included in order to account for the highly adversarial nature of cyber operations. Because both the attacker and the defender operate often with very limited awareness of each other's actions, situational awareness of both should be modeled. One of the workshop's presentations pointed out the importance of situational awareness in achieving impact on the opponent's mission.

In many cases, however, both the defender and the attacker are best modeled not as individual human cognitive actors, but rather as organizations. Organizational modeling is studied by a community of researchers in the political science field that is distinct from the community of cognitive modelers. It would be worth exploring how that community might help solving the MIA problem.

## 2.6 Model Construction

The current practice of constructing models for MIA is almost entirely manual in nature. As such, model construction is very time consuming, expensive, and difficult to document, inspect, and validate. Maintenance of such models – also manual – is also expensive. Quantitative characterization of dependencies between, for example, business functions and supporting technical assets, is largely a matter of asking the presumed subject matter experts (SMEs) for a number, such as a conditional probability. The guessing of such numbers of SMEs is expensive and the verity of numbers is doubtful.

Still, manual construction of models for MIA problems is feasible, even if expensive. For example, Analyzing Mission Impacts of Cyber Actions (AMICA) (reported in one of the workshop's papers) is a comprehensive MIA modeling and simulation tool with a fully implemented military "business" model. It relies on manually crafted models.

Some tools exist that allow essentially manual yet computer-aided construction of business models. The widely available Business Process Management (BPM) tools fall into this category. This may be important because, in general, systems (and networks/infrastructure) are better understood than missions (business), and the availability of BPM tools can help close this gap.

Ideally, however, we would like to see the bulk of MIA models constructed automatically, perhaps by observing a business process and its cyber defense operations, and automatically learning or inferring a model. One of the workshop's presentations described an approach where significant part of a model was automatically derived from the observed network flows.

Modeling of missions can be assisted by decomposing missions hierarchically into sub-missions, e.g., strategic into tactical. Although this aids in simplifying and better understanding of mission, it adds the problem of modeling complex interactions between elements of the decomposed structure. As in any model, it is important to balance granularity and determine the minimum necessary level of fidelity.

## 2.7 Data Requirements

Modelling and simulation techniques depend upon availability of large empirical datasets, and MIA problems are no exception. All aspects of models we have discussed depend on availability of data that can be used to create model elements. Data are needed to create, validate, and maintain such models. Therefore, it is necessary to establish a rigorous, comprehensive program of automated collection and, in some cases, manual maintenance of the data.

Examples of the required data include the topology of the information technology (IT) and communication systems under consideration; the cyber vulnerabilities of the software and hardware involved; the business processes of the organization; the complex interdependencies between missions, services, and infrastructure; and behaviors of cyber actors.

## 3. Conclusion

### 3.1 Selected Observations on the Current State of R&D in MIA (Outcome of Brainstorming Sessions)

- A degree of maturity has been achieved in models of dependencies between cyber assets and physical assets.

- Linking attack graphs and missions has been demonstrated.

- Examples exist of models that use game theory and control theory for the purposes of MIA.

- Attempts have been made to model the cyber adversary.

- Comprehensive examples exist where certain business processes were modeled with sufficient fidelity to be used for MIA.

- There is understanding and some examples of how to quantify vulnerability, risk and resiliency in order to assess mission impact.

- Relevant models for risk assessment do exist.

- Generally, human inputs are manually collected to populate models with parameters.

## 3.2  Selected Suggestions for the Future of R&D in MIA (Outcome of Brainstorming Sessions)

### 3.2.1  Operations

- Methods and techniques to reduce the time to detect the cyberattacks or compromises so that mission impact can be minimized.

- Novel architectures that minimize the mission impact by design

- Defense tactics, techniques, and procedures (TTPs) can be explicitly optimized to control the extent of mission impact.

- Intuitive situational awareness to support commander decision making and automated execution of actions

### 3.2.2  Formalisms

- Formal languages for MIA that capture mission, attacker, defense, security architecture, etc.

- Formal models for situation awareness, such as hierarchical concurrent probabilistic finite state machine (FSM)

- Formal language for decision making integrated with formal cyber language

### 3.2.3  Automation

- Automated infrastructure model creation

- Extract model parameters automatically from logs

- Automated population of situational awareness models

- Automated support to model validation, and to training and exercises

- Appropriate mix of automated and human response to mission impact

### 3.2.4 Data

- Collect ground-truth data on attacker and defender behaviors

- Remote cyber sensing for collecting attacker data

- Open source, vendors and users are valuable providers of data

- Better characterization and use of synthetic data and sanitized real data

## 3.3 Recommendations for a Follow-on Activity

A suitable follow-on to this workshop would be an exploratory team (ET) activity that focuses on a study of potential approaches and likely barriers to applying a model-driven paradigm to a broad range of cyber defense activities, including but not limited to, mission impact assessment.

The ET could be titled, tentatively, "The Model-Driven Paradigms for Integrated Approaches to Cyber Security." The ET would explore the possibility of using a multi-purpose, integrated system of models for guiding a broad range of cybersecurity operations: vulnerability analysis, intrusion prevention, intrusion detection, analysis, forensics, attribution, mission impact assessment, and recovery. The team will investigate whether such a paradigm can be expected to provide meaningful, actionable information about cyber activities and the resulting impacts that have not been seen before or do not match prior experiences and patterns.

The workshop already identified a number of research and development (R&D) challenges associated with model-driven approaches to MIA. Therefore, it will be the ET's charter to answer a number of questions, such as the following:

- How likely these challenges be overcome in the near- and mid-term?

- Are there applications of the model-driven paradigm that are more likely to prove fruitful in near-term for the MIA problem?

- What can be learned and adopted from the Canadian Automated Computer Network Defence (ARMOUR) demonstrator and European Union PANOPTESEC prototype, which explore a related model-based approach?

- What are ways to populate and validate models in an affordable fashion?

- Is the model-driven paradigm defeated by ever-growing diversity and diffusion of IT infrastructures, such as Internet of Things (IoT)?

- What commercial tools are emerging that can support the model-driven paradigm?

- Can any standards can be leveraged to enable model-driven R&D to be conducted in an interoperable manner? If not, should candidates for standards be defined within an STO activity?

## 4. References

1. Kott A, Stoyanov N, Baykal N, Moller A, Sawilla R, Jain P, Lange M, Vidu C. Proceedings of the NATO IST-128 Workshop Assessing Mission Impact of Cyberattacks. Dec 2015. Adelphi, MD: Army Research Laboratory (US). Report No.: ARL-RP-0562.

INTENTIONALLY LEFT BLANK.

# Appendix A. The Announcement of the Workshop IST-128

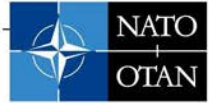
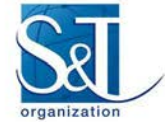

SCIENCE AND TECHNOLOGY ORGANIZATION

INFORMATION SYSTEMS TECHNOLOGY PANEL

**WORKSHOP ANNOUNCEMENT**
**CALL FOR CONTRIBUTIONS**

Cyber Attack Detection, Forensics and Attribution for

Assessment of Mission Impact

IST-128-RWS-019

to be held in

Istanbul, Turkey

15 - 17 June 2015

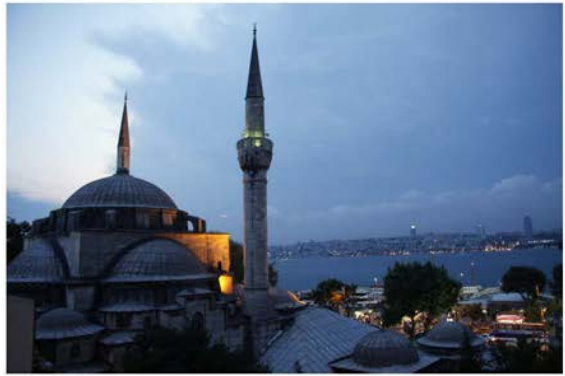

IF YOU WISH TO PRESENT A PAPER, SUBMIT BEFORE 1 April 2015

ENROL BEFORE 1 June 2015

The Workshop is NATO UNCLASSIFIED
OPEN to Partnership for Peace (PfP/EAPC) Nations

BP 25 - F-92201 Neuilly-sur-Seine Cedex - France
Tel: +33 (0)1 55 61 22 00 - Fax: +33 (0)1 55 61 22 99 - E-Mail : mailbox@cso.nato.int

# The NATO Science and Technology Organization

Science & Technology (S&T) in the NATO context is defined as the selective and rigorous generation and application of state-of-the-art, validated knowledge for defence and security purposes. S&T activities embrace scientific research, technology development, transition, application and field-testing, experimentation and a range of related scientific activities that include systems engineering, operational research and analysis, synthesis, integration and validation of knowledge derived through the scientific method.

In NATO, S&T is addressed using different business models, namely a collaborative business model where NATO provides a forum where NATO Nations and partner Nations elect to use their national resources to define, conduct and promote cooperative research and information exchange, and secondly an in-house delivery business model where S&T activities are conducted in a NATO dedicated executive body, having its own personnel, capabilities and infrastructure.

The mission of the NATO Science & Technology Organization (STO) is to help position the Nations' and NATO's S&T investments as a strategic enabler of the knowledge and technology advantage for the defence and security posture of NATO Nations and partner Nations, by conducting and promoting S&T activities that augment and leverage the capabilities and programmes of the Alliance, of the NATO Nations and the partner Nations, in support of NATO's objectives, and contributing to NATO's ability to enable and influence security and defence related capability development and threat mitigation in NATO Nations and partner Nations, in accordance with NATO policies.

The total spectrum of this collaborative effort is addressed by six Technical Panels who manage a wide range of scientific research activities, a Group specialising in modelling and simulation, plus a Committee dedicated to supporting the information management needs of the organization.

- AVT     Applied Vehicle Technology Panel
- HFM     Human Factors and Medicine Panel
- IST     Information Systems Technology Panel
- NMSG    NATO Modelling and Simulation Group
- SAS     System Analysis and Studies Panel
- SCI     Systems Concepts and Integration Panel
- SET     Sensors and Electronics Technology Panel

These Panels and Group are the power-house of the collaborative model and are made up of national representatives as well as recognised world-class scientists, engineers and information specialists. In addition to providing critical technical oversight, they also provide a communication link to military users and other NATO bodies.

The scientific and technological work is carried out by Technical Teams, created under one or more of these eight bodies, for specific research activities which have a defined duration. These research activities can take a variety of forms, including Task Groups, Workshops, Symposia, Specialists' Meetings, Lecture Series and Technical Courses

**The Information Systems Technology (IST) Panel**

The mission of the Information Systems Technology (IST) Panel is to implement, on behalf of the S&T Board, the STO Mission with respect to Information Systems Technology. The advancement and exchange of techniques and technologies to provide timely, affordable, dependable, secure and relevant information to war fighters, planners and strategists, as well as enabling technologies for modelling, simulation, and training are the focus of this Panel.

The **Information Systems Technology Panel (IST)** covers the fields of:

a) Architecture & Intelligent Information Systems (A2IS)
b) Communications and Networks (COM)
c) Information Warfare and Assurance (IWA)

## PROGRAMME COMMITTEE

### Programme Chair

**Dr. Alexander KOTT**
U.S. Army Research Laboratory
United States
Email: alexander.kott1.civ@mail.mil

### COMMITTEE MEMBERS

Mr. Hüseyin TİRLİ
TÜBİTAK BİLGEM
Cyber Security Institute
Email: huseyin.tirli@tubitak.gov.tr
Turkey

Mr. Edmund N. MOXON
DSTL
Email: enmoxon@dstl.gov.uk
United Kingdom

Dr. Reginald SAWILLA
NATO-NCI Agency
Email: reginald.sawilla@ncia.nato.int
The Netherlands

Assoc. Prof. Dr. Maya BOZHILOVA
Defence Institute "Prof. Tsvetan Lazarov"
Email: m.bozhilova@di.mod.bg
Bulgaria

Dr. Dennis McCALLAM
Northrop Grumman
Email: dennis.mccallam@ngc.com
United States

Mr. Oğuzhan TOPGÜL
TÜBİTAK BİLGEM
Cyber Security Institute
Turkey

Col. Assoc. Prof. Dr. Nikolai STOIANOV
Defence Institute "Prof. Tsvetan Lazarov"
Email: n.stoianov@di.mod.bg
Bulgaria

Dr. Chris WILLIAMS
Dstl
Email: cwilliams@dstl.gov.uk
United Kingdom

Mr. Alfred MØLLER
Danish Defence Acquisition and Logistics Organization
Email: avm@mil.dk
Denmark

## BACKGROUND

The success of a military mission is highly dependent on the Communications and Information Systems (CIS) hat support the mission, and on their use in the cyber battle space. The inexorably growing dependency on computational information processing for weapons, intelligence, communication, and logistics systems continues to increase the vulnerability of missions to various cyber threats.

Attacks on CIS systems or other cyber incidents degrade or disrupt the usage of CIS systems, and the resulting mission capability, performance, and completion. Such incidents are expected to increase in frequency and sophistication. Therefore there is a need to address the technology and procedures to characterize the impact of cyber attacks on the mission. Such an impact analysis must necessarily include a broad range of cyber analysis activities: detect attacks in a mission-supporting manner, assess damages relevant to the mission, investigate impacts on mission elements, recover from attacks in order to continue missions to the maximum extent possible, and decide on how to respond to cyber attacks in a manner that maximizes mission success.

Additionally, forensics methods and tools are necessary to determine key facts relevant to assessing mission impact. Such tools are used for evidence collection, analysis of the attack, identification of the attacker, understanding the attack, damage assessment, and attribution of attackers. Dependent on the mission and the type of an attack - there may be different degrees of relative importance and resources attached to attack detection, continuity of the military mission, damage assessment, evidence collection, attribution, and other activities. Usage of related methods, procedures, tools or technology should depend largely on mission.

## WORKSHOP OBJECTIVES:

The workshop will focus on identifying practice and research challenges, gaps and approaches – current and future -- to assessment of mission impact due to a cyber attack. Because such an assessment is inseparable from, and impossible without, attack detection, forensics and attribution, the workshop will explore how these activities and related technologies and methods should support the assessment of mission impact.

An example of a complex challenge to be explored by this workshop is achieving the right balance between computational, communication and information resources. Such a balance must be maintained between resources required for ongoing military operations and mission success, resources used for attack detection, battle damage assessment, and investigation - including forensics methods - and resources used to identify origin of attacker in order to determine the attack response, while optimizing the likelihood of mission success. An example of a potential opportunity that the workshop may explore is the question of whether principles known from the traditional military battle damage identification and assessment can be utilized in the cyber domain.

By examining such complex issues, the workshop will formulate a coherent structure of key research and development challenges and technology gaps, as well as recommendations for most promising research and development approaches to closing the gaps.

## WORKSHOP TOPICS:

All example topics focus on the goal of assessment of mission impact. They include but are not limited to:

- Analysis and modeling of mission and mission dependencies of CIS assets;
- Prediction of mission impact, including cascading impacts;
- Quantification and qualification of predicted mission risks and impacts;
- Quantification of the criticality of assets in accordance with mission dependencies;
- Methods and techniques for assessing mission dependencies and cyber risks;

- Incident analysis from mission impact perspective - methods, tools and technology;
- Mission-focused attack detection with prioritization for mission needs, including early warning;
- Advanced data analysis tools for characterizing attackers tools used in the incident;
- Automated damage assessment;
- Mission-focused forensics of information, computers and networks ;
- Automation of mission-focused forensics triage;
- Tools and methods for visualization of damage and the impact on mission dependencies;
- Correlation and fusion of damage and evidence data;
- Mission impact focused attribution and trace back;
- Current and future trends, including potential for real-time or large scale forensics and other analysis that characterizes impact on a particular mission. Emerged/emerging "disruptive" technology developments;
- Metrics for mission impact assessments;
- Use of simulation, e.g., event (re)construction methods and tools, and simulation of impact on mission, such as dependencies propagation.

## WORKSHOP FORMAT AND ORGANIZATION

The workshop will bring together military and civilian cyber security researchers, technologists, and practitioners. The workshop provides an excellent opportunity to increase participants' insight in civil as well as military cyber security problem space, and to influence the future research in cyber security, especially as it relates to mission impact assessment and related detection, forensics and attribution. The workshop will comprise a series of topical sessions – see Attachment 1. Each session will include several presentations of papers – some of which will be full papers and others will be shorter position papers – and a discussion open to all participants. The Programme Committee will utilize both the papers and the oral discussions at the workshop to formulate the final report of the workshop, including a set of recommendations.

## CALL FOR PAPERS

Although submission of a paper is not required for participating in the workshop, it is encouraged. The Programme Committee invites two types of papers: full papers and position papers. A full paper is a technical paper (between 2500-5000 words) that presents results of a novel research and is subject to evaluation criteria of a typical technical conference. A position paper is short (between 500-1500 words) and reflects the views of the author – usually a discussion of research challenges, gaps and approaches -- without necessarily presenting a supporting research. Papers must address one or more of the aforementioned topics and focus on assessment of mission impact due to a cyber attack.

All (NATO UNCLASSIFIED-Releasable to PfP) papers must be submitted and sent by e-mail to the Workshop Chairman (alexander.kott1.civ@mail.mil) and to the Chairman's Assistant (ana.a.santiesteban.civ@mail.mil ) by the deadline set in the schedule (see below). US authors and non-US Citizen affiliated with a US organization, please see Attachment 2.

The paper must include the following information, in the beginning:
- IST-128 Workshop on Cyber Attack Detection, Forensics and Attribution for Assessment of Mission Impact
- TITLE OF THE PAPER
- Name of the principle Author, followed by the names of the Co-Author(s) if any, and then Company/Affiliation, complete mailing addresses, telephone, fax and e-mail addresses

-5-

It is the responsibility of each contributor to fulfil the publication release and clearance requirements of his/her organization/company and country to obtain clearance of papers as needed. An official clearance is mandatory in the United States and there may also be a requirement in other countries to obtain clearance for unclassified papers. For further information, authors may contact any of the Programme Committee Members listed in this document or their National STO Coordinator. Please allow sufficient time for the clearance to be issued before deadline. In this case, the NATO classification for the Workshop has been declared as NATO UNCLASSIFIED – Releasable to Partner for Peace (PfP) nations.

**US Authors**: Authors from the United States must comply with US procedures.
*(Refer to the Instructions in Attachment 2)*

The Programme Committee will select a number of papers that are considered suitable for presentation at the Workshop. Authors will be notified by the date indicated in the schedule whether or not their papers are selected. Authors of selected papers will also be provided with information in the Instructions for Authors, which contains detailed instructions for the final formatting, presentation, transmission, etc. of papers.

The time allowed for each presenter of a full paper is 20 minutes, for a position paper - 10 minutes. Equipment will be available for PowerPoint presentations. Paper presentation times will be given in the Programme Announcement included with General Information Package. All papers accepted for presentation at the workshop will appear in the Workshop's Report and published electronically on the CSO Website.

Please note that the authors of papers selected for presentation will not be financially supported by this organization. You are fully responsible for your own hotel and travel.

### Schedule

| | |
|---|---|
| 1 April | Deadline for submission of your paper (not required for attendance but encouraged) |
| 15 May | Author is notified whether the paper has been accepted |
| 1 June | Deadline for enrolment (whether you present a paper or not) |

## GENERAL INFORMATION

### Classification

All material and discussion in this workshop will be unclassified.

### Participation and Enrolment

You can attend and participate in the workshop even if you do not present a paper. However, enrolment is required in order to attend the workshop. Whether you present a paper or not, you must enrol for the workshop on the CSO website (www.cso.nato.int) before 1st June 2015. The enrolment web page will become available on March 15. We encourage you to enrol early. Attendance will be limited to a number of people to be determined by the Programme Committee.

### Language

Presentations and discussions will be in English.

### Workshop site, lodging and social programme

The workshop will be held in Istanbul, Turkey.

There is no workshop registration fee.

Attendees and accompanying persons will be responsible for their own accommodation arrangements and any travel expenses.

-6-

Approved for public release; distribution unlimited.

Once you have enrolled on the CSO website and your enrolment has been validated, you will automatically receive a General Information Package (GIP), giving you further details about the meeting site, the hotels and other general information.

Any questions on the technical aspects of the scientific programme or the participation process should be addressed to the Workshop Chair.

Questions on the administrative aspects of this Workshop or requests for further information on STO activities should be addressed to the IST Panel Office:

(Interim) IST Panel Executive
Mr. Philippe SOÈTE
E-mail: philippe.soete@cso.nato.int
Tel: +33 (0)1 5561 2280

IST Panel Assistant
Mrs. Ayşegül APAYDIN
E-mail: aysegul.apaydin@cso.nato.int
Tel: +33 (0)1 5561 2282

Science & Technology Organization/Collaboration Support Office (CSO)
Information Systems Technology (IST) Panel
BP 25, 922001 Neuilly sur Seine, France

Questions on the local arrangement and facilities should be addressed to Mr. Hüseyin TİRLİ (huseyin.tirli@tubitak.gov.tr)

The Programme and Organization of the Workshop

The Tentative Programme and Organization of the Workshop

**MONDAY, 15 June 2015**
09h00 – 12h00      Session 1:  Mission Impact Assessment and Attack Detection

In this session we explore the complex relations between the assessment of mission impact and the various aspects of detection of malicious cyber activities. Understanding of potential nature of mission impact, and priorities associated with mission impact, can guide the process of detection. Conversely, the information gleaned in the process of detection informs assessment of the mission impact and helps its automation.

13h00 – 16h00      Session 2:  Mission Impact Assessment and Forensics

Cyber forensics is a key process that yields insights into the nature of the impact that a cyber incident induces on the mission. As such, the forensics itself should be informed by the mission structure and dynamics. Here we explore mission-focused forensics of information, computers and networks; automation of mission-focused forensics triage; characterizing attackers tools; real-time or large scale forensics analysis that characterizes impact on a particular mission.

**TUESDAY, 16 June 2015**
09h00 – 12h00      Session 3:  Mission Impact Assessment and Attribution

The challenge of attribution and trace-back for cyber incidents continues to grow in importance and complexity. Mission impact assessment can provide clues regarding the intent and the identity of potential perpetrators of the cyber attack. At the same time, any information about the attacker can help guide and focus the impact assessment. We also explore correlation and fusion of related data.

13h00 – 16h00      Session 4:  Modeling, Simulations and Visualization for Mission Impact Assessment

In this session we explore analysis and modeling of mission and mission dependencies of CIS assets; use of event reconstruction methods and tools, and simulation of impact on mission, such as dependencies propagation. Also of interest are tools and methods for visualization of damage and the impact on mission; and metrics for mission impact assessments.

**WEDNESDAY, 17 June 2015**
09h00 – 12h00      Session 5:  Opportunities, Priorities and Recommendations

In a structured brainstorming process, this session will integrate and analyze the overall results and findings of the workshop; it will yield key clusters of technical gaps and barriers, promising approaches to overcoming the gaps, and prioritization of directions in research and development of methods, technologies and tools for mission impact assessment. Recommendations for follow-on NATO IST activities, if any, will be formulated as well.

**SPECIAL NOTICE FOR US AUTHORS AND
NON-US CITIZENS AFFILIATED WITH US ORGANIZATIONS**

Papers from the U.S. must be sent **ONLY** to the following P.O.C.:

NATO STO U.S. National Coordinator
OASD (R&E)/International Technology Programs
4800 Mark Center Drive, Suite 17D08
Alexandria, VA 22350-3600
E-mail: david.r.uribe.ctr@mail.mil   or   usnatcor@osd.mil
Tel: +1 571 372 6539 / 6538
Fax: +1 571 372 6471

PLEASE NOTE THE FOLLOWING:

All U.S. Authors must submit one electronic copy to this P.O.C. by **1 APRIL 2015**

All US Authors must include the following statement in a covering letter:
- The work described in this paper is cleared for presentation to NATO audiences (i.e., Approved for public release)
- The paper is technically correct
- If work is sponsored by a government agency, identify the organization and attest that the organization is aware of submission
- The paper is NATO/PfP Unclassified; and
- The paper does not violate any proprietary rights.

NOTE:
1. Only complete packages (paper plus all items listed above) will be accepted by the US P.O.C.
2. After review and approval, the US POC will forward all US papers with the Details of Authors Form to the Panel Assistant. All US papers must be received directly from the US POC. US papers will not be accepted directly from authors.
3. In the event there are any questions or concerns with these requirements, U.S. authors are encouraged to contact the US POC as early as possible. Delays in meeting POC deadlines will impact the timely submission of your paper.

# Appendix B. The Final Program of the Workshop IST-128

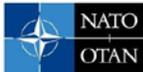

# SCIENCE & TECHNOLOGY ORGANIZATION
## COLLABORATION SUPPORT OFFICE

## WORKSHOP

### Cyber Attack Detection, Forensics and Attribution for Assessment of Mission Impact

**IST-128 / RWS-019**

organised by the
**Information Systems and Technology Panel**

to be held in
**Istanbul, Turkey**

Monday 15 June 2015 - Wednesday 17 June 2015

This Workshop is NATO Unclassified
open to PfP and MD nations

**Latest Enrolment Dates**

| | |
|---|---|
| NATO Nations | Monday, 1 June 2015 |
| Others | Friday, 29 May 2015 |

**Enrol Online at:**
http://www.cso.nato.int

Once your enrolment is validated, you will receive a General Information Package (GIP) giving you further necessary details about the meeting.

If you are unable to enrol via the internet, please contact the

IST Panel Assistant at:
aysegul.apaydin@cso.nato.int

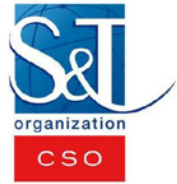

### Background
Information Systems Technology Panel (IST) is one of the seven Panels whose role it is to implement, on behalf of the Science & Technology Board, the STO Mission with respect to Information Systems Technology. The advancement and exchange of techniques and technologies to provide timely, affordable, dependable, secure and relevant information to war fighters, planners and strategists, as well as enabling technologies for modelling, simulation, and training are the focus of this Panel. The Information Systems Technology Panel covers the fields of Information Warfare and Assurance, Information and Knowledge Management, Communications and Networks and Architecture and Enabling Technologies.

### Theme - Objectives – Topics
The success of a military mission is highly dependent on the Communications and Information Systems (CIS) hat support the mission, and on their use in the cyber battle space. The inexorably growing dependency on computational information processing for weapons, intelligence, communication, and logistics systems continues to increase the vulnerability of missions to various cyber threats.

Attacks on CIS systems or other cyber incidents degrade or disrupt the usage of CIS systems, and the resulting mission capability, performance, and completion. Such incidents are expected to increase in frequency and sophistication. Therefore there is a need to address the technology and procedures to characterize the impact of cyber-attacks on the mission. Such an impact analysis must necessarily include a broad range of cyber analysis activities: detect attacks in a mission-supporting manner, assess damages relevant to the mission, investigate impacts on mission elements, recover from attacks in order to continue missions to the maximum extent possible, and decide on how to respond to cyber- attacks in a manner that maximizes mission success. Additionally, forensics methods and tools are necessary to determine key facts relevant to assessing mission impact. Such tools are used for evidence collection, analysis of the attack, identification of the attacker, understanding the attack, damage assessment, and attribution of attackers. Dependent on the mission and the type of an attack - there may be different degrees of relative importance and resources attached to attack detection, continuity of the military mission, damage assessment, evidence collection, attribution, and other activities. Usage of related methods, procedures, tools or technology should depend largely on mission

This Workshop will focus on identifying practice and research challenges, gaps and approaches – current and future -- to assessment of mission impact due to a cyber attack. Because such an assessment is inseparable from, and impossible without, attack detection, forensics and attribution, the workshop will explore how these activities and related technologies and methods should support the assessment of mission impact.

### Collaboration Support Office - Point of Contact
Mrs. Aysegül APAYDIN   Fax: +33 (1) 5561 9626
IST Panel Assistant    Email: aysegul.apaydin@cso.nato.int
NATO STO – CSO
Tel: +33 (1) 5561 2282

### WORKSHOP TOPICS:
All example topics focus on the goal of assessment of mission impact. They include but are not limited to
- Analysis and modeling of mission and mission dependencies of CIS assets;
- Prediction of mission impact, including cascading impacts;
- Quantification and qualification of predicted mission risks and impacts;
- Quantification of the criticality of assets in accordance with mission dependencies;
- Methods and techniques for assessing mission dependencies and cyber risks;

### Programme Committee
#### Programme Chair

**Dr. Alexander KOTT**
US Army Research Laboratory
USA
Email: alexander.kott1.civ@mail.mil

#### Members

**Mr. Hüseyin TİRLİ**
TÜBITAK BILGEM
Cyber Security Institute
Turkey
Email: huseyin.tirli@tubitak.gov.tr

**Mr. Oğuzhan TOPGÜL**
TÜBITAK BILGEM
Cyber Security Institute
Turkey
Email: oguzhan.topgul@tubitak.gov.tr

**Prof. Bob MADAHAR**
DSTL
United Kingdom
Email: bkmadahar@mail.dstl.gov.uk

**Col.Assoc.Prof.Dr. Nikolai STOIANOV**
Defence Institute "Prof. Tsyetan Lazarov"
Bulgaria
Email: n.stoianov@di.mod.bg

**Assoc.Prof.Dr. Maya BOZHILOVA**
Defence Institute "Prof. Tsyetan Lazarov"
Bulgaria
Email: m.bozhilova@di.mod.bg

Dr. Dennis McCALLAM
Northrop Grumman
United States
Email: dennis.mccallam@ngc.com

Mr. Alfred MØLLER
Danish Defence Acquisition and Logistics Organization
Denmark
Email: avm@mil.dk

Dr. Reginald SAWILLA
NATO NCI Agency
Netherlands
Email: reginald.sawilla@ncia.nato.int

Dr. Chris WILLIAMS
DSTL
United Kingdom
Email: cwilliams@dstl.gov.uk

## Science and Technology Organization in NATO

In NATO, Science & Technology (S&T) is defined as the selective and rigorous generation and application of state-of-the-art, validated knowledge for defence and security purposes. S&T activities embrace scientific research, technology development, transition, application and field-testing, experimentation and a range of related scientific activities that include systems engineering, operational research and analysis, synthesis, integration and validation of knowledge derived through the scientific method.

In NATO, S&T is addressed using different business models:

- The Collaborative business model where NATO provides a forum where NATO Nations and partner Nations elect to use their national resources to define, conduct and promote cooperative research and information exchange.
- The In-House delivery business model where S&T activities are conducted in a NATO dedicated executive body, having its own personnel, capabilities and infrastructure.

## The Science and Technology Organization - STO

The mission of the NATO STO is to help position the Nations' and NATO's S&T investments as a strategic enabler of the knowledge and technology advantage for the defence and security posture of NATO Nations and partner Nations, by:

- Conducting and promoting S&T activities that augment and leverage the capabilities and programmes of the Alliance, of the NATO Nations and the partner Nations, in support of NATO's objectives;
- Contributing to NATO's ability to enable and influence security- and defence-related capability development and threat mitigation in NATO Nations and partner Nations, in accordance with NATO policies;
- Supporting decision-making in the NATO Nations and NATO.

## Acknowledgements

We wish to thank ITÜ, the Istanbul Technical University, for hosting this Workshop in Istanbul.

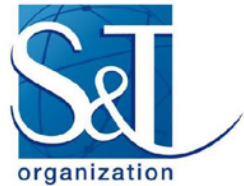

http://www.sto.nato.int

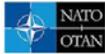

# IST-128 Cyber Attack Detection, Forensics and Attribution for Assessment of Mission Impact

## Programme

### Monday 15 June 2015

**Registration**
- 08:30 Registration

**Introductory Session**
- 09:00 Welcome Speech 1
  by Prof. Dr. Mehmet KARACA, Rector of Istanbul Technical University (ITU), TUR

  Welcome Speech 2
  by Prof.Dr. Ahmet Arif ERGIN, President of TUBITAK BILGEM, TUR

  Overview of STO IST Panel and Related Initiatives
  by Dr. John McLEAN, IST Panel Chairman, USA

  Introduction to Workshop
  by Dr. Alexander KOTT, Workshop Chairman, USA
- 10:00 BREAK

**SESSION 1 - Understanding the Adversary in Order to Understand the Mission Impact**
- 10:30  1  Cyber Weapons Development and Testing - Detection, Attribution, Assessment and Deterrence: An Important Challenge to the R&D Community
  by Samuel VISNER, ISF Internationl, USA
- 11:00  2  Mission Impact and the Role of Behavioral Science
  by Cleotilde GONZALEZ, Carnegie Mellon University, USA
- 11:30  PANEL DISCUSSION
- 12:00  LUNCH

**SESSION 2 - Assessing the Mission Impact from Observable Signals and Events**
- 13:00  3  Probabilistic Mission Imapct Assessment Based on Widespread Local Events
  by Alexander MOTZEK, Ralf MÖLLER, Mona LANGE, University of Lübeck, DEU, Samuel DUBUS, ALTACEL Lucent, FRA
- 13:30  4  Software Correlation for Malware Characterization
  by Philippe CHARLAND, Defence R&D Canada, CAN
- 14:00  5  Estimating Attack Intent and Mission Impact from Detection Signals
  by Patrick McDANIEL, Rober WALLS, Pennsylvania State University, USA
- 14:20  BREAK
- 14:40  PANEL DISCUSSION
- 16:00  End of Day

### Tuesday 16 June 2015

**SESSION 3 - Modeling Systems and Missions for Mission Impact Assessment**
- 09:00  6  Mission Impact Assessment in Power Grids
  by Mona LANGE, Ralf MÖLLER, University of Lübeck, DEU, Marina KROTOFIL, European Network for Cyber Security
- 09:30  7  Sensory Channel Threats to Military CPS and IoT Assets
  by Selcuk ULUAGAC, Florida International University, USA
- 10:00  8  Mission Assurance as a Function of Scale
  by Pierre TREPAGNIER, Alexia SCHULZ, MIT Lincoln Laboratory, USA
- 10:30  BREAK
- 11:00  PANEL DISCUSSION
- 12:00  LUNCH

**SESSION 4 - Modeling and Simulations of Mission Impact**
- 13:00  9  Cyber-Attack as a Contest Game
  by Alexander ALEXEEV, Odessa State Ecological University, UKR, Kerry KRUTILLA, Indiana University, USA
- 13:30  10  Cyber Risk Analysis of CIS-Dependent Missions: A Modeler's Perspective on Preparing for Detecting and Responding to Cyber Attacks for Assessment of Mission Impact
  by Matthew HENRY, David ZARET, Ryan CARR, Daniel GORDON, Johns Hopkins University, USA
- 14:00  11  Analyzing Mission Impacts of Cyber Actions (AMICA)
  by Steven NOEL, Jackson LUDWIG, Prem JAIN, Dale JOHNSON, Roshan K. THOMAS, Jenny McFARLAND, Ben KING, MITRE Corporation, USA, Seth WEBSTER, Brady TELLO, MIT Lincoln Laboratory, USA
- 14:30  BREAK
- 15:00  PANEL DISCUSSION
- 16:00  End of Day

### Wednesday 17 June 2015

**SESSION 5 - Opportunities, Priorities and Recommendations**
- 09:00  In a structured brainstorming process, this session will integrate and analyze the overall results and findings of the Workshop. It will yield key clusters of technical gaps and barriers, promising approaches to overcoming the gaps, and prioritization of directions in research and development of methods, technologies and tools for mission impact assessment. Recommendations for follow-on NATO IST activities, if any, will be formulated as well.
- 12:00  CLOSING OF EVENT

## List of Symbols, Abbreviations, and Acronyms

| | |
|---|---|
| ACT-R | Adaptive Control of Thought—Rational |
| AMICA | Analyzing Mission Impacts of Cyber Actions |
| ARMOUR | Automated Computer Network Defence |
| BPM | Business Process Management |
| CIA | confidentiality, integrity, and availability |
| CISs | communications and information systems |
| ET | exploratory team |
| FSM | finite state machine |
| IoT | Internet of Things |
| IST | Information Systems and Technology |
| IT | information technology |
| MD | Mediterranean Dialogue |
| MIA | mission impact assessment |
| NATO | North Atlantic Treaty Organization |
| OCO | observability, controllability, and operability |
| PtP | Partnership for Peace |
| R&D | research and development |
| S&T | science and technology |
| SME | subject matter expert |
| STO | Science and Technology Organization |
| TTPs | tactics, techniques, and procedures |